\documentclass[a4paper]{jpconf}
\usepackage{graphicx}
\usepackage{amsmath, amsfonts}
\usepackage{mathrsfs}
\usepackage{sidecap}
\usepackage{lipsum}
\usepackage{amsthm}
\usepackage{lineno}
\begin{document}
\title{Constraining 2HDM+S model through W-boson mass measurements}

\author{Anza-Tshilidzi Mulaudzi$^{a}$, Mukesh Kumar$^{a}$, Ashok Goyal$^c$, Bruce Mellado$^{a,b}$}

\address{$^{a}$School of Physics and Institute for Collider Particle Physics, University of the Witwatersrand, 1 Jan Smuts Avenue, Johannesburg, 2050, South Africa}
\address{$^{b}$iThemba LABS, National Research Foundation, PO Box 722, Somerset West, 7129, South Africa}
\address{$^{c}$Department of Physics \& Astrophysics, University of Delhi, Delhi, 110 007, India}

\ead{anza-tshilidzi.mulaudzi@cern.ch,mukesh.kumar@cern.ch,agoyal45@yahoo.com,\\bmellado@mail.cern.ch}

\begin{abstract}
Following a discussion on $W$-boson mass observed at the CDF and ATLAS, we explore the parameter space allowed in the 2HDM+$S$ model. Further, the model parameter space is constrained through vector-like leptons via muon $g-2$ measurements. We show our results for additional scalar mass fixed to $m_S \approx 95$ and $150$~GeV keeping the standard Higgs-boson mass at 125~GeV in all four types of 2HDM+$S$ model. The chosen mass of the singlet scalar is motivated by the excesses seen at the CMS and ATLAS data in proton-proton collisions at the Large Hadron Collider.
\end{abstract}

\section{Introduction}
In recent times, there has been growing concern arising about the precise measurements of the $W$-boson mass. These evaluations appear to reveal noteworthy disparities in comparison to the anticipated $W$-boson mass within the Standard Model (SM), denoted as $M_W^{\rm SM} = 80.357 \pm 0.006$~GeV. These concerns come after the recent measurement from the CDF collaboration where the measurement exhibits a notable deviation of 7$\sigma$ from the SM prediction. The deviation of $M_W^{\rm CDF}$ from the SM prediction has been attributed to the contributions from the new physics beyond the SM (BSM). An attractive approach to explain this deviation is through the contributions to Peskin-Takeuchi oblique~\cite{Peskin:1990zt} parameters in new physics models~\cite{Kumar:2013yoa}. So our purpose is to explore the parameter space of a BSM which is motivated by the multi-lepton anomalies and also using the benchmark points motivated by the excesses seen at CMS at 95~GeV and also the combinations from the ATLAS and CMS datasets of different channels at 152~GeV, we want to see which parameter space can be fitted for the $W$ mass measurement. The anomalies seen in the multi-leptonic channels in CMS~\cite{CMS:2012qbp,CMS:2013btf} and ATLAS~\cite{ATLAS:2012yve} experiments in the SM Higgs boson decay have been attributed to the possible existence of low-mass scalar boson resonances typically in the mass range of 90~GeV – 200~GeV. The excesses observed in the $\gamma \gamma$, $\tau^+ \tau^-$, $b \bar{b}$ and $W^+W^-$ channels can also be attributed to arise from similar mass resonances~\cite{Sabatta:2019nfg}. A Two-Higgs doublet model of Type II Yukawa is used to study them and the presumed dominant decays are $H\to Sh$, $SS$, which leads to the search for new scalar resonances decays $S\to \gamma\gamma$, $Z\gamma$ in association with missing transverse energy, light- and $b$-jets. Subsequently, we would like to show the muon $g-2$ with the contributions from vector-like leptons (VLL), so that we could see what the $W$ boson mass from the VLL contributions can be fitted to the SM prediction.  

\section{Model}
The formalism of the 2HDM$+S$ is a 2HDM 
extended by $\Phi_{S}$ which is a real singlet field and the potential is defined by~\cite{vonBuddenbrock:2016rmr,Muhlleitner:2016mzt,Ivanov:2017dad}:
\begin{align}        
V(\Phi_{1},\Phi_{2},\Phi_{S}) =&\, m_{11}^{2}|\Phi_{1}|^{2} - m_{22}^{2}|\Phi_{2}|^{2} - m_{12}^{2}(\Phi_{1}^{\dagger}\Phi_{2}+h.c.) +
        \frac{\lambda_{1}}{2}(\Phi_{1}^{\dagger}\Phi_{1})^{2}+\frac{\lambda_{2}}{2}(\Phi_{2}^{\dagger}\Phi_{2})^{2} \notag\\
        &\,+
        \lambda_{3}(\Phi_{1}^{\dagger}\Phi_{1})(\Phi_{2}^{\dagger}\Phi_{2}) + 
        \lambda_{4}(\Phi_{1}^{\dagger}\Phi_{2})(\Phi_{2}^{\dagger}\Phi_{1}) +
        \frac{\lambda_{5}}{2}[(\Phi_{1}^{\dagger}\Phi_{2})^{2}+ {\rm h.c.}] \notag\\
        &\,+ 
        \frac{1}{2}m_{S}^{2}\Phi_{S}^{2} + \frac{\lambda_{6}}{2}\Phi_{S}^{4} + \frac{\lambda_{7}}{2}(\Phi_{1}^{\dagger}\Phi_{1})\Phi_{S}^{2} + \frac{\lambda_{8}}{2}(\Phi_{2}^{\dagger}\Phi_{2})\Phi_{S}^{2}.
    \label{THDMSpot}
\end{align}
The 2HDM$+S$ model involves the Higgs doublets $\Phi_{1}$ and $\Phi_{2}$ associated with $SU(2)_{L}$ symmetry, and the singlet $\Phi_S$ contributes to the potential as shown in the final line of equation~\ref{THDMSpot}. When dealing with theories containing multiple Higgs doublets, the problem of tree-level flavour-changing neutral currents (FCNC) often emerges. To address this issue, a clever strategy is employed where each Higgs doublet couples to quarks which is achieved through $\mathbb{Z}_{2}$ symmetry. This symmetry is softly broken by the $m_{12}^{2}$ term. The tree-level FCNC interactions are eliminated by extending the $\mathbb{Z}_{2}$ symmetry to the Yukawa sector and in order to achieve this, we require 
\begin{equation}
    \Phi_{1}\longrightarrow \Phi_{1}, \Phi_{2}\longrightarrow -\Phi_{2}, \Phi_{S}\longrightarrow \Phi_{S}.
\end{equation}

In this model, we explore a scenario in which the real singlet field $\Phi_{S}$ obtains a $vev$ while preserving the $\mathbb{Z}_{2}'$ symmetry. To achieve this, we introduce a soft breaking of the $\mathbb{Z}_{2}$ symmetry by setting $m_{12}^{2}\neq 0$ in the 2HDM$+S$ potential. Consequently, we assume that $\lambda_{i}$ are real, resulting in a model devoid of explicit CP violation. The presence of an additional real singlet field alters the CP-even neutral sector of the 2HDM$+S$, distinguishing it from the original 2HDM configuration. Therefore, we have
\begin{equation}
    \begin{array}{c}
         M_{CP-even}^{2} \\
    \end{array} = 
    \left(\begin{array}{ccc}
        \lambda_{1}^{2}c_{\beta}^{2}v^{2} + t_{\beta}m_{12}^{2} & \lambda_{345}c_{\beta}s_{\beta}v^{2} - m_{12}^{2} & \lambda_{7}c_{\beta}vv_{S} \\
        \lambda_{345}c_{\beta}s_{\beta}v^{2} - m_{12}^{2} & \lambda_{2}^{2}s_{\beta}^{2}v^{2} + m_{12}^{2}/t_{\beta} & \lambda_{8}s_{\beta}vv_{S} \\
        \lambda_{7}c_{\beta}vv_{S} & \lambda_{8}s_{\beta}vv_{S} & \lambda_{6}v^{2}_{S} 
    \end{array}\right)
\end{equation}

The 2HDM$+S$ model is governed by a total of 12 independent real parameters. To ensure physical significance, we select as many parameters as possible. By employing minimization conditions, we can interchange $m_{11}^{2}$, $m_{22}^{2}$, and $m_{S}^{2}$ with the Standard Model vev ($v$), $t_{\beta}$, and $v_{S}$, respectively. The quartic couplings are expressed in terms of the physical masses and mixing angles. We use the following set of input parameters considering the Type II 2HDM:
\begin{equation}
    t_{\beta}, v, v_{S},\alpha_{1}, \alpha_{2}, \alpha_{3}, m_{H_{1,2,3}}, m_{A}, m_{H^{\pm}}, m_{12}^{2}
\end{equation}

\section{Theoretical and experimental constraints}
\label{theory}
The theoretical and experimental constraints are applied through the use of the $\textsc{ScannerS}$ program. $\textsc{ScannerS}$ applies experimental constraints at 95\% confidence level (CL), which is $\approx 2\sigma$. Therefore, the points that satisfy (or obey) the theoretical framework are returned by \textsc{ScannerS} but they are not rejected by observations at $\approx$ 95\% CL. These constraints include vacuum stability, boundedness from below, perturbative unitarity, electroweak precision, flavour physics, Higgs searches and measurements and electroweak phase transition. The constraints in \textsc{ScannerS} can be implemented at three different levels, \textbf{skip}, \textbf{ignore} and \textbf{apply}. The weakest level, \textbf{skip}, ignores the particular constraint and performs no calculations that are associated with it, thus no parameter points will be tested against it. In the intermediate level, \textbf{ignore}, all calculations are performed and the obtained results are kept. The strongest level, \textbf{apply}, only accepts points that satisfy the particular constraint. For a detailed explanation of each of these constraints, we can take a look at Refs.~\cite{Biekotter:2022abc,Muhlleitner:2016mzt,djouadi1998hdecay,Butterworth:2010ym,engeln2019n2hdecay}


\section{The $M_{W}$ calculation} 
To ensure that the model complies with the electroweak precision observables (EWPOs), the constraints can be represented in terms of Peskin-Takueshi parameters, $U$, $S$, and $T$~\cite{Haller:2018nnx,Takeuchi:1992bu}. As is true for extended Higgs sectors, new physics contributions to these parameters can be substantial if they mostly occur through gauge boson self-energies. From $\textsc{ScannerS}$ we integrate the one-loop corrections to the Peskin-Takueshi parameters for the Type II Yukawa 2HDM, therefore, the $W$-boson mass will be calculated as a function of the $S$, $T$, and $U$ parameters, defined by~\cite{Grimus:2008nb}
\begin{equation}
    M^{2}_{W} = \left(M^{\rm SM}_W\right)^2 \left(1+\frac{s^{2}_{w}}{c^{2}_{w}-s^{2}_{w}}\Delta r'\right),
    \label{mw}
\end{equation}

with 
\begin{equation}
    \Delta r' = \frac{\alpha}{s^{2}_{w}}\left(-\frac{1}{2}S+c^{2}_{w}T +\frac{c^{2}_{w}-s^{2}_{w}}{4s^{2}_{w}}U\right),
    \label{deltaR}
\end{equation}

and 
\begin{equation}
    \sin^{2}\theta_{\rm eff} = \sin^{2}\theta_{\rm eff}^{\rm SM} - \alpha \frac{c^{2}_{w}s^{2}_{w}}{4(c^{2}_{w}-s^{2}_{w})}T,
\end{equation}
where $M_{W}^{\rm SM} = 80.537$~GeV, $\sin^{2}\theta_{\rm eff}^{\rm SM} = 0.231532$, $s_{w} = \sqrt{1-\left(\frac{M_{W}^{\rm SM}}{M_{Z}}\right)^{2}}$, $c_{w} = \sqrt{1 - s_{w}^{2}}$ and $M_{Z} = 91.1876$~GeV.

\section{Performing a fitting procedure for the enhancements observed at 95 and 150~GeV}
A $\chi^{2}$ analysis is done to explain the $\gamma \gamma$, $\tau^{+}\tau^{-}$, $b \Bar{b}$ excesses that were observed and also verify whether the parameter space of the 2HDM$+S$ can predict an upward shift in the $W-$boson mass. The experimental signal strengths for the three different channels were observed to be 

\begin{align}
    &\mu^{exp}_{\gamma\gamma} \pm \Delta\mu^{exp}_{\gamma\gamma} = 0.27^{+0.10}_{-0.09} ~\cite{Biekotter:2023oen}\\
    &\mu^{exp}_{bb} \pm \Delta\mu^{exp}_{bb} = 0.117 \pm 0.057~\cite{OPAL:2002ifx}\\
    &\mu^{exp}_{\tau\tau} \pm \Delta\mu^{exp}_{\tau\tau} = 1.2 \pm 0.5~\cite{CMS:2022goy}
\end{align}

An additional $WW$ channel is incorporated into the $\chi^2$ with a signal strength of $\mu^{exp}_{WW} \pm \Delta \mu^{exp}_{WW} = 14.6 \pm 6.8$~\cite{Coloretti:2023wng}
The signal strengths are defined by
We define the signal strengths as
\begin{equation}
    \mu_{XX} = \frac{\sigma^{BSM}(gg\rightarrow H)\cdot BR^{SM}(H\rightarrow XX)}{\sigma^{SM}(gg\rightarrow H(m_{BSM}))\cdot BR^{SM}(H(m_{BSM})\rightarrow XX)}
\end{equation}
The uncertainties from experiments are reported as $1\sigma$ variations. The values $\mu_{\gamma\gamma}$, $\mu_{bb}$, $\mu_{\tau\tau}$ and, $\mu_{WW}$ representing the experimentally observed signal strengths were acquired through parameter scans using $\textsc{ScannerS}$. We use $\textsc{ScannerS}$to compute the BSM Higgs-like scalars which leads to the formulation of $\chi^2$ contributions for each observed excess and it is expressed as 

\begin{equation}
    \chi^{2}_{\gamma\gamma,\tau^{+}\tau^{-},b \Bar{b},WW} = \frac{(\mu_{\gamma\gamma,b\Bar{b},\tau\tau,WW}-\mu^{exp}_{\gamma\gamma,b\Bar{b},\tau\tau,WW})^{2}}{(\Delta\mu^{exp}_{\gamma\gamma,b\Bar{b},\tau\tau,WW})^{2}} 
\end{equation}
 
In order to evaluate the comprehensive representation of the three excesses, along with an extra excess in the $WW$ channel (where signal rates were derived from $\textsc{ScannerS}$ and observed results from experiments were obtained from~\cite{Coloretti:2023wng}), we establish the cumulative $\chi^{2}$ contribution as follows
\begin{equation}
    \chi^{2}_{\gamma\gamma,\tau^{+}\tau^{-},b \Bar{b}, WW} = \chi^{2}_{\gamma \gamma}+\chi^{2}_{\tau^{+}\tau^{-}} + \chi^{2}_{b \Bar{b}} + \chi^{2}_{WW},
\end{equation}

In the case of three measurements being considered independently, this criterion corresponds to the condition that
\begin{equation}
    \chi^{2}_{\gamma\gamma,\tau^{+}\tau^{-},b \Bar{b},WW}  \leq 4
\end{equation}
We perform a random scan in the 2HDM$+S$ for a 95~GeV scalar and 152~GeV scalars. The parameter ranges are defined as 
\begin{equation}
    \begin{array}{ccc}
        m_{h_{1}}= 94~{\rm GeV} - 98~{\rm GeV}, & m_{h_{2}}=125.09~{\rm GeV}, & 200~{\rm GeV}\leq m_{h_{3}} \leq 500~{\rm GeV},\\
        400~{\rm GeV}\leq m_{A} \leq 1000~{\rm GeV}, & -\pi/2 \leq \alpha_{1,2,3} \leq \pi/2, & 400~{\rm GeV}\leq m_{h^{\pm}} \leq 1500~{\rm GeV}, \\
        0.5 \leq tan\beta \leq 4.0, & 0 \leq m^{2}_{12} \leq 5\times10^{5}, & 1~{\rm GeV}\leq v_{S} \leq 1500~{\rm GeV}
        \label{Scan96}
    \end{array}
\end{equation}
Subsequently, the parameter ranges for a 152~GeV scalar are given by 
\begin{equation}
    \begin{array}{ccc}
        m_{h_{1}}=125.09~{\rm GeV}, & m_{h_{2}}=148~{\rm GeV} - 153~{\rm GeV}, & 250~{\rm GeV}\leq m_{h_{3}} \leq 800~{\rm GeV},\\
        400~{\rm GeV}\leq m_{A} \leq 1500~{\rm GeV}, & -\pi/2 \leq \alpha_{1,2,3} \leq \pi/2, & 400~{\rm GeV}\leq m_{h^{\pm}} \leq 1500~{\rm GeV}, \\
        0.5 \leq tan\beta \leq 1.0, & 0 \leq m^{2}_{12} \leq 5\times10^{5}, & 1~{\rm GeV}\leq v_{S} \leq 1500~{\rm GeV}
        \label{Scan150}
    \end{array}
\end{equation}

It's important to highlight that this exploration primarily targets a parameter region where $S_{95}$, representing $m_{h_1}$, exhibits sufficiently strong interactions with gauge bosons and fermions, enabling it to account for the observed and computed excesses in Higgs searches at 95~GeV. For the provided parameters, we use $\textsc{ScannerS}$ to randomly generate points and provide the signal strengths for the four channels. $\textsc{ScannerS}$ incorporates the theoretical and experimental constraints discussed in Sect.\ref{theory}. We only considered points that fall within the 2$\sigma$ region of the new ATLAS measurement,
\begin{equation}
    \chi^{2}_{M_{W}^{\rm{World Avg.}}} = \frac{(M_{W}^{\rm {2HDM}+S} - M_{W}^{\rm{World Avg.}})^{2}}{(\Delta M_{W}^{\rm{World Avg.}})^{2}} \leq 4.
\end{equation}
with $\Delta M_{W}^{\rm {World Avg.}} = 0.015$~GeV.

\begin{figure}
\centering
{\includegraphics[width=.48\textwidth]{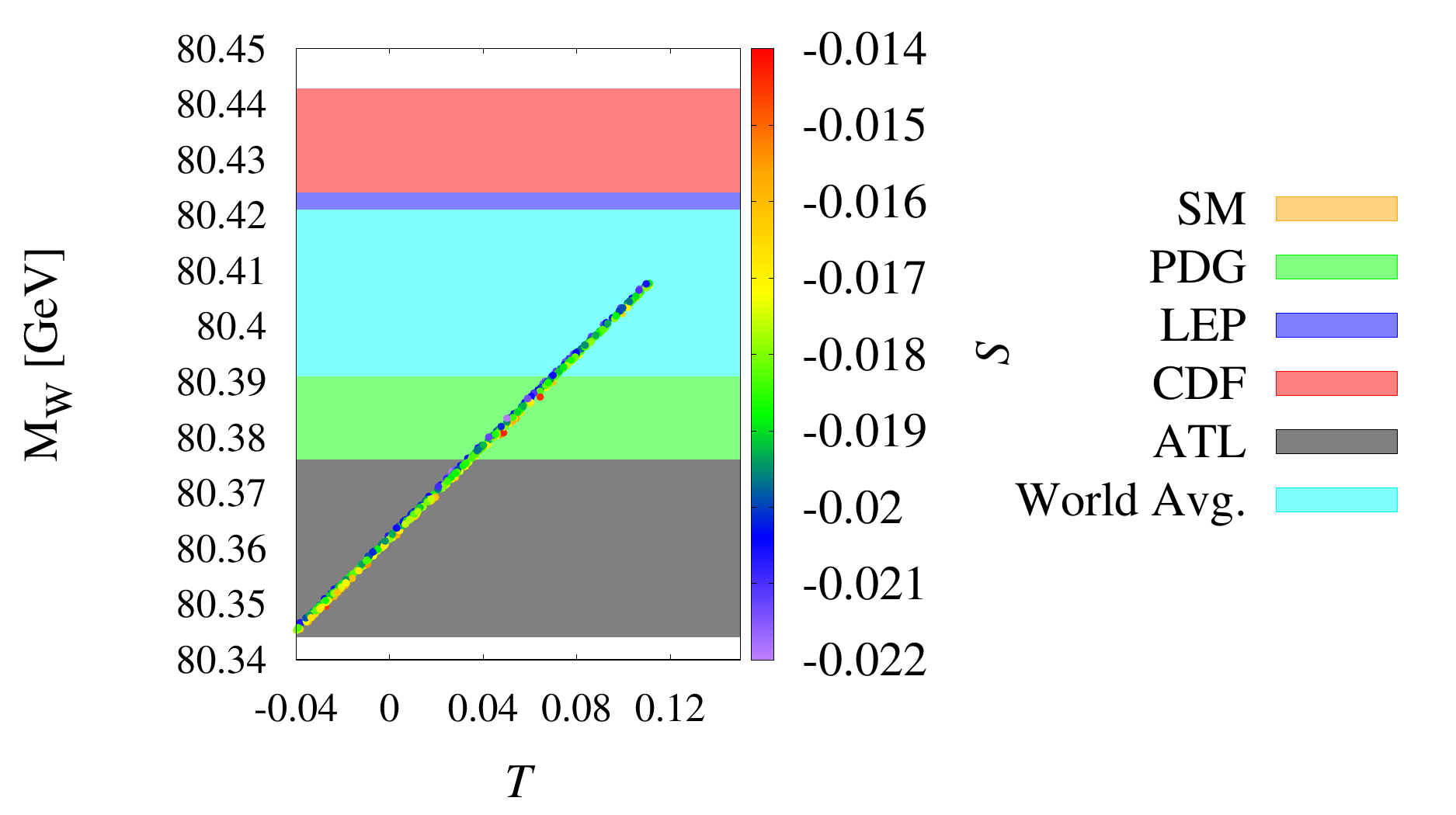}}
{\includegraphics[width=.48\textwidth]{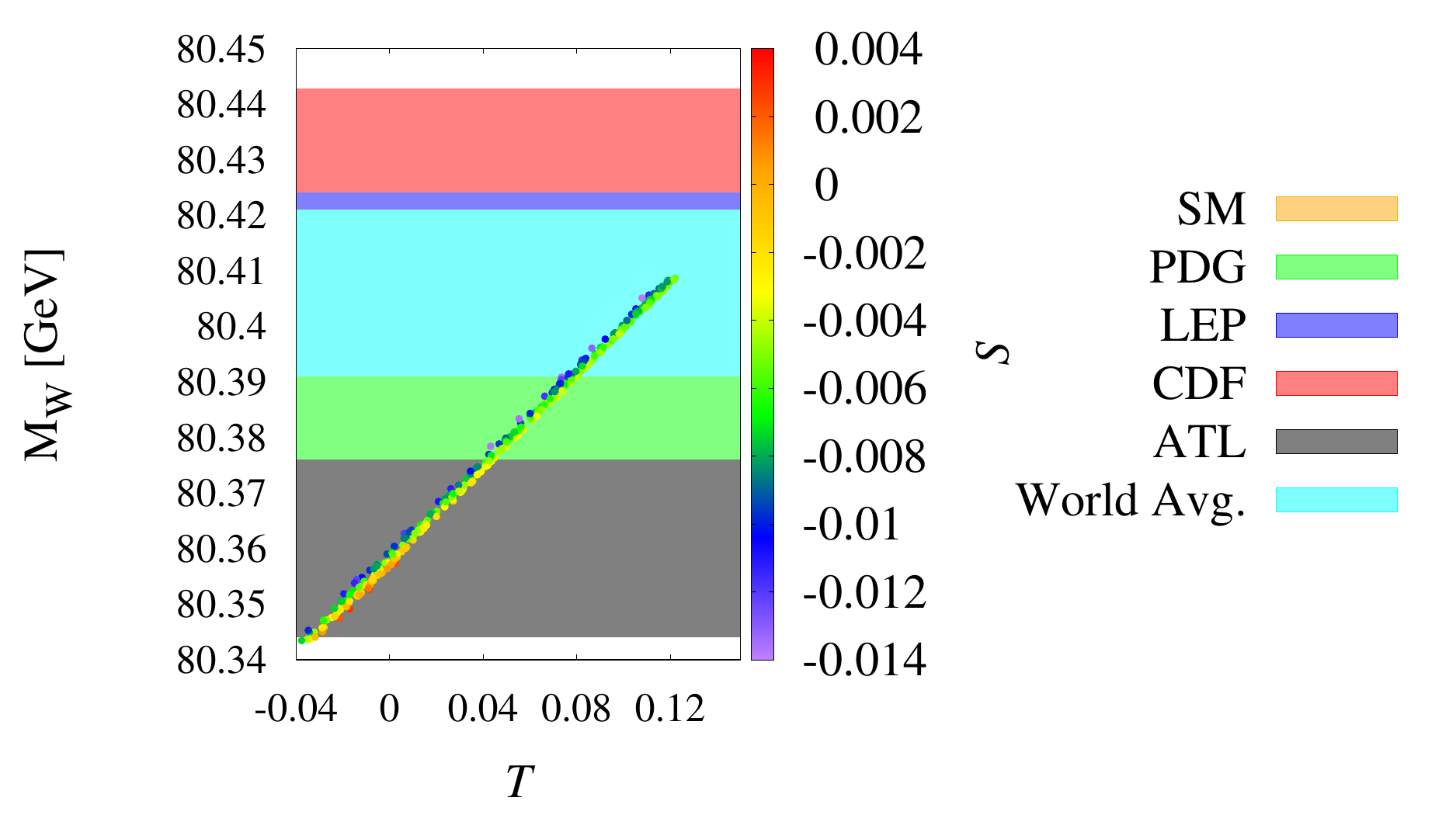}}
\caption{The $M_{W}$ prediction in the 2HDM$+S$ parameter space is depicted here for the 95~GeV~(left) and 152~GeV~(right) scalars. It is shown in conjunction with the Peskin-Takueshi parameters $T$ and $S$. The bands represent the value of the $M_{W}$ value as measured by each experiment, the world average and, the SM prediction. \label{fig:MwScan}}
\end{figure}

\begin{figure}
\centering
{\includegraphics[width=.48\textwidth]{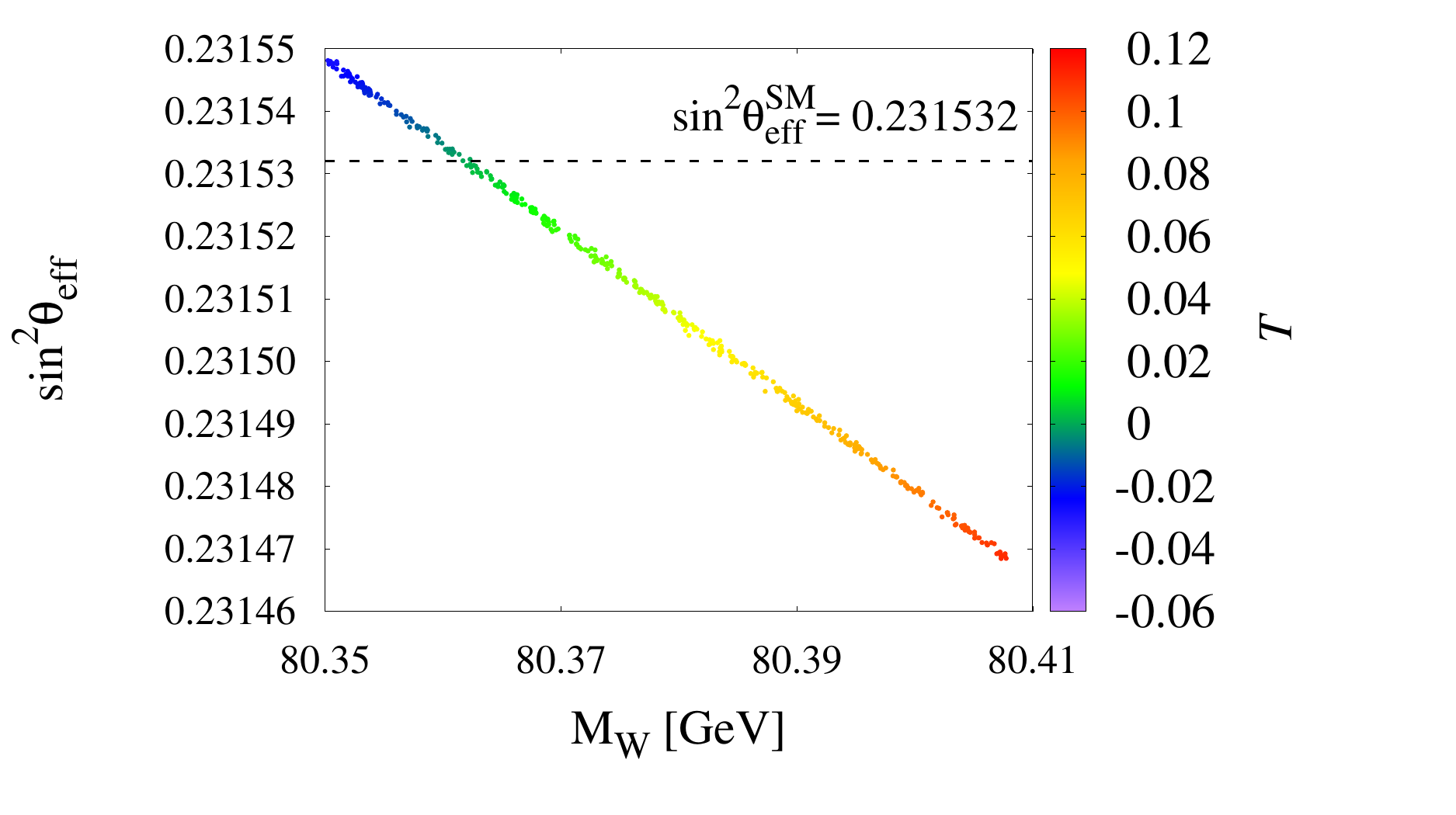}}
{\includegraphics[width=.48\textwidth]{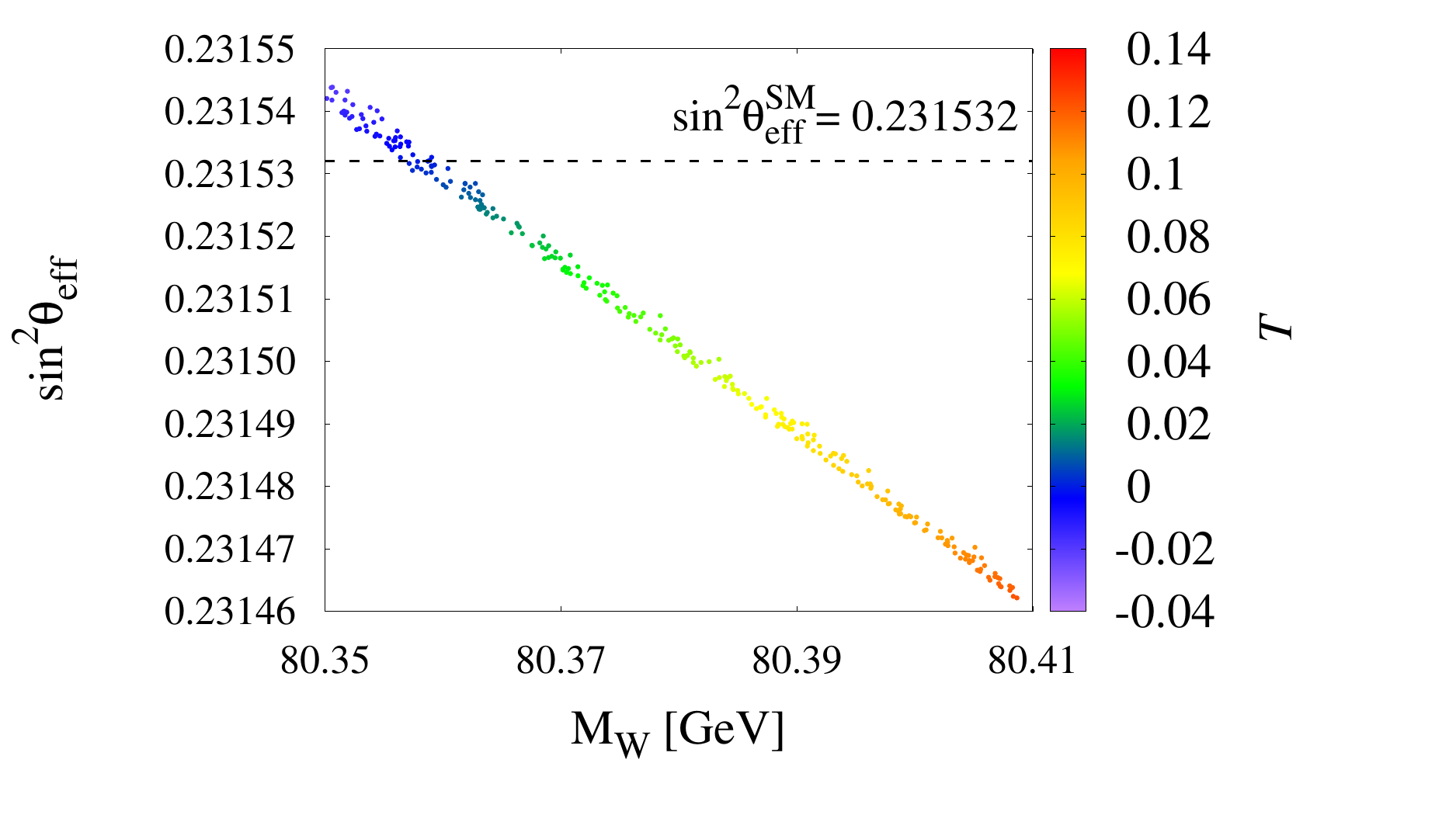}}
\caption{The plots illustrate the projections of the $W$ boson mass and the effective weak mixing angle in the 2HDM$+S$ model for 95~GeV (left) and 152~GeV (right) scalars. These projections are depicted as a function of the Peskin-Takueshi parameter, $T$. The SM prediction of $\sin^{2}\theta_{\rm eff}^{\rm SM}$ is represented as a dashed line.\label{fig:SineffScan}}
\end{figure}

In Figure~\ref{fig:MwScan}, we see that the contoured points agree with measurements from PDG, ATLAS and the world average. Subsequently, our parameter space does not meet the measurements from the CDF and LEP experiments. It can be seen that the value of $M_{W}$ includes the predictions from the SM, PDG and, the ATLAS experiment for the ranges of $S$ and $T$. This remains the case for both 95 and 152~GeV scalars. 

Figure~\ref{fig:SineffScan} shows the predictions for $M_{W}$ and $\sin^{2}\theta_{\rm eff}$ in the 2HDM$+S$ for all points with $\chi^{2}_{M_{W}^{\rm{2HDM}+S}} \leq 4$. The results show that with the chosen parameter space of the 2HDM$+S$, the $W-$ boson mass satisfies the predictions of the different experiments but it does not satisfy the CDF and LEP measurements of $M_{W}$. We also see that the parameter points that $M_W$ fit the new ATLAS measurement of the $W-$boson mass feature are also comparable $\sin^{2}\theta_{\rm eff}$ results compared to the SM prediction.

\section{Conclusions}
The CDF reported a $7\sigma$ deviation of the $W-$ boson mass compared to the SM prediction. However, the ATLAS experiment recently reported an updated measurement of the mass of the $W$ boson and this measurement favours the SM prediction. This paper delves into the feasibility of accommodating different experimental $M_W$ measurements within the parameter space of a Type II 2HDM model, augmented with an additional Singlet Scalar (2HDM+$S$). Our focus is directed towards the Type II Yukawa structure of the 2HDM, chosen for its suitability in explaining excesses observed in the quest for Higgs-like scalars in the mass range of 130 GeV to 160 GeV. Through our analysis, we establish that the 2HDM+$S$ model yields a prediction for $M_W$ that concurs with a substantial body of experimental measurements. Moreover, this prediction harmonizes effectively with collider excesses observed at 95 GeV and 152 GeV. This congruence extends to aligning with diverse theoretical and experimental constraints imposed on the model's parameters. Additionally, our study reveals that the parameter space we consider can aptly accommodate the effective weak mixing angle in the context of the new $M_W$ measurement, reinforcing its compatibility with SM predictions

Furthermore, this prediction demonstrates a notable agreement with collider excesses detected at both 95~GeV and 152~GeV. This consistency extends to adhering to various theoretical and experimental constraints imposed on the model's parameters. Additionally, our investigation indicates that the parameter space under examination can appropriately accommodate the effective weak mixing angle in light of the new measurement of $M_{W}$, further affirming its compatibility with Standard Model predictions. Looking ahead, we intend to explore the inclusion of Beyond the Standard Model (BSM) fermions as one-loop contributions to $\Delta a_\mu$. Specifically, our interest lies in the incorporation of singly charged SM singlet vector-like leptonic fermions as presented in the Lagrangian of Ref.~\cite{Freitas:2014pua}. The interplay between muon and vector-like singlet leptonic fermion $f'$ mixing can introduce corrections to the muon mass.

\end{document}